\documentclass[12pt]{iopart}
\usepackage{latexsym}
\usepackage[pdftex]{graphicx}
\usepackage{bm}
\usepackage{epsfig}

\begin{document}

\title[]{Solution of the phase problem for coherent scattering from a disordered system of identical particles}

\author{R P Kurta$^1$, R Dronyak$^1$\footnote{deceased}, M Altarelli$^2$, E Weckert$^1$, I A Vartanyants$^1$,$^3$}

\address{$^1$ Deutsches Elektronen-Synchrotron DESY, Notkestra\ss e 85, D-22607 Hamburg, Germany}
\address{$^2$ European X-ray Free-Electron Laser Facility, Notkestra\ss e 85, D-22607, Hamburg, Germany}
\address{$^3$ National Research Nuclear University,``MEPhI'', 115409, Moscow, Russia}
\ead{ivan.vartaniants@desy.de}

\date{\today}

\begin{abstract}
While the implementation of single particle coherent diffraction imaging for non-crystalline particles is complicated by current limitations in photon flux, hit rate, and sample delivery a concept of many-particle coherent diffraction imaging offers an alternative way to overcome these difficulties.
Here we present a direct, non-iterative approach for the recovery of the diffraction pattern corresponding to a single particle using coherent x-ray data collected from a two-dimensional (2D) disordered system of identical particles, that does not require {\it a priori} information about the particles and can be applied to a general case of particles without symmetry.
The reconstructed single particle diffraction pattern can be directly used in common iterative phase retrieval algorithms to recover the structure of the particle.

\end{abstract}

\pacs{61.05.cf, 87.59.-e, 61.46.Df}

\maketitle

\section{Introduction}

It was recently realized \cite{Wochner1,Altarelli,Kurta} that analysis of diffraction patterns from disordered systems based on intensity cross-correlation functions (CCFs) can provide information on the local symmetry of these systems.
A variety of disordered systems, for example, colloids and molecules in solution, liquids, or atomic clusters in the gas phase can be studied by this approach. It becomes especially attractive with the availability of x-ray free-electron lasers (FELs) \cite{Emma, SCSS, Altarelli2}. Essentially, scattering from an ensemble of identical particles can provide, in principle, the same structural information as single particle coherent imaging experiments \cite{Gaffney} that are presently limited in the achievable resolution \cite{Hajdu}. Clearly, a large number of particles in the native environment can scatter up to a higher resolution, at the same photon fluences, comparing to experiments on single particles injected into the FEL beam.
However, it is still a challenge to recover the structure of individual particles composing a system using the CCF formalism.
In the pioneering work of Kam \cite{Kam, Kam1} it was proposed to determine the structure of a single particle using scattered intensity from many identical particles in solution.
However, this approach, based on spherical harmonics expansion of the scattered amplitudes, was not fully explored until now.
Recently, it has been revised theoretically \cite{Saldin, Saldin1, Saldin4} and experimentally \cite{Saldin2}. The possibility to recover the structure of individual particle was demonstrated in systems in two-dimensions (2D) \cite{Saldin,Saldin1,Saldin2} and in three-dimensions (3D) \cite{Saldin4} using additional {\it a priori} knowledge on the symmetry of the particles. Unfortunately, these approaches, based on optimization routines \cite{Saldin,Saldin4} and iterative techniques \cite{Saldin1, Saldin2}, do not guarantee the uniqueness of the recovered structure unless strong constraints are applied. This all leads to a high demand in finding direct, non-iterative approaches for recovering the structure of an individual particle in the system.

In this paper we further develop Kam's ideas and propose an approach enabling unambiguous reconstruction of single-particle
intensity distributions using algebraic formalism of measured two- and three-point CCFs without additional constraints. Once the single-particle intensity is obtained, conventional phase retrieval algorithms \cite{Fienup, Elser} can recover the projected electron density of a single particle. Our approach is developed for 2D systems of particles which is of particular interest for studies of membrane proteins. It can be also used to study 3D systems provided that particles can be aligned with respect to the incoming x-ray beam direction.

\section{Basic equations \label{sec:basic}}

\begin{figure}[!htbp]
\centering
\includegraphics[width=0.7\textwidth]{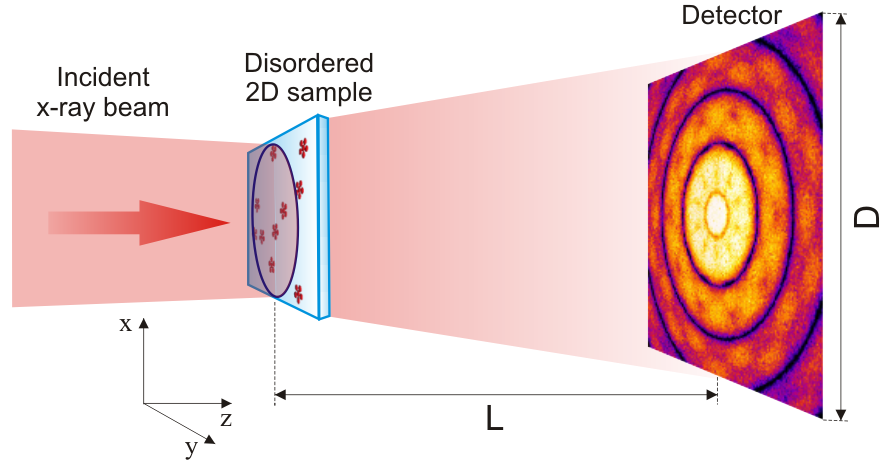}
\caption{\label{Fig:ExpGeometr}(Color online) Geometry of the diffraction experiment. The incident x-ray beam coherently illuminates a 2D disordered sample and produces a diffraction pattern on a detector. The direction of the incident beam is defined along the z axis of the coordinate system.}
\end{figure}

\begin{figure}[!htbp]
\centering
\includegraphics[width=0.6\textwidth]{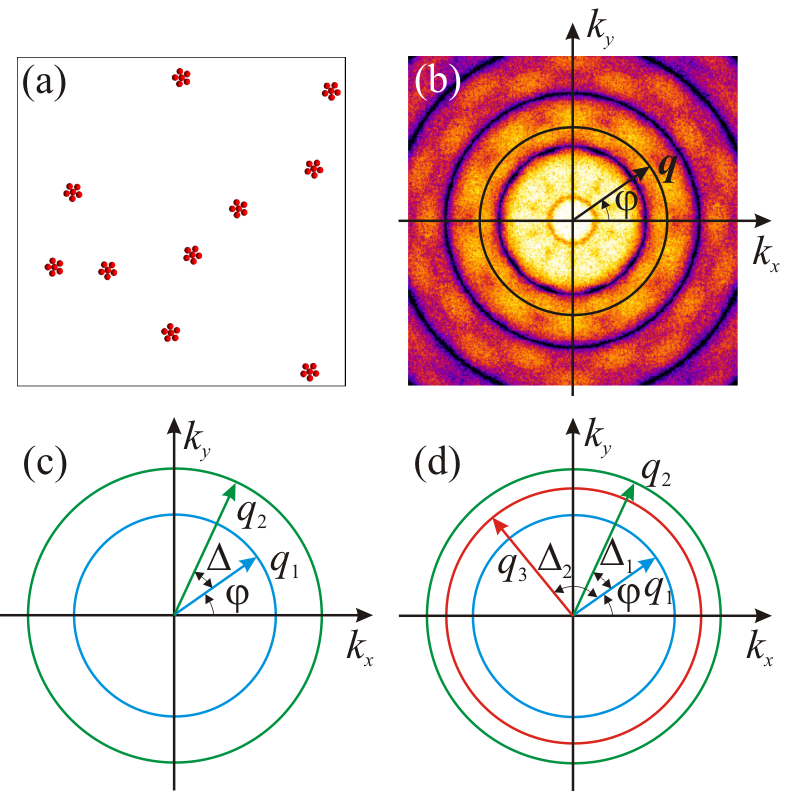}
\caption{\label{Fig:CCFScheme}(Color online)
(a) 2D disordered system composed of $N=10$ clusters in random positions and orientations (b) Instantaneous scattered intensity $I(q,\varphi)$ as a function of the angular position $\phi$ on the resolution ring $q$. (c),(d) Definition of the momentum transfer vectors in the derivation of the two-point $C(q_{1},q_{2},\Delta)$ (c), and the three-point CCFs $C(q_{1},q_{2},q_{3},\Delta_{1},\Delta_{2})$ (d) .}
\end{figure}

We consider a scattering experiment in transmission geometry [see figure~\ref{Fig:ExpGeometr}], where the direction of the incoming x-ray beam is perpendicular to the 2D sample plane, and simulate a set of diffraction patterns.
Our sample consists of an arbitrary small number $N$ of identical, spatially disordered particles [see figure~\ref{Fig:CCFScheme}(a)]. Also, we assume a uniform distribution of orientations of particles in the system.
First, for simplicity, we consider a model where the total scattered intensity $I(\bi{q})$ is represented as an {\it incoherent} sum of intensities $I_{\psi_{i}}(\bi{q})$ corresponding to individual particles in the system,
\begin{equation}
I(\bi{q})=\sum\limits_{i=1}^{N} I_{\psi_{i}}(\bi{q}),\label{Iincoh1}
\end{equation}
where $\bi{q}$ is the momentum transfer vector and $\psi_{i}$ is the orientation of the $i$-th particle. This model is a good approximation for dilute systems when
the mean distance between particles is much larger than their size \cite{Altarelli,Kurta}. In this approximation we neglect the inter-particle correlations due to coherent interference of scattered amplitudes from individual particles. Later, in simulations, we generalize our approach to the case of {\it coherent} scattering from a system of particles in the presence of Poisson noise and demonstrate the applicability of our approach.

In the frame of kinematical scattering, the intensity $I_{\psi_{0}}(\bi{q})$ scattered from a single particle in some reference orientation $\psi_{0}$ is related to the electron density of the particle $\rho_{\psi_{0}}(\bi{r})$ by the following relation
\begin{equation}
I_{\psi_{0}}(\bi{q})=\left| \int \rho_{\psi_{0}}(\bi{r})\exp(\rmi\bi{qr}) \rmd\bi{r} \right|^{2}.\label{Iro}
\end{equation}
Once the single-particle intensity $I_{\psi_{0}}(\bi{q})$ is determined, conventional phase retrieval algorithms \cite{Fienup, Elser} can recover the projected electron density $\rho_{\psi_{0}}(\bi{r})$ of a single particle.
Our goal is to determine the scattering pattern of a single particle $I_{\psi_{0}}(\bi{q})$ using a large number of diffraction patterns $I(\bi{q})$ corresponding to different realizations of the system.

The intensity $I_{\psi_{0}}(\bi{q})$ scattered  from a single particle can also be described in a polar coordinate system $\bi{q}=(q,\varphi)$ as a function of the angular position $(0<\varphi \leq 2 \pi)$ on the resolution ring $q$ [see figure~\ref{Fig:CCFScheme}(b)]. For each resolution ring $q$, one can represent the intensity function $I_{\psi_{0}}(q,\varphi)$  as a Fourier series expansion,
\begin{equation}
I_{\psi_{0}}(q,\varphi)=\sum\limits_{n=-\infty}^{\infty} I_{q,\psi_{0}}^{n}\exp(\rmi n\varphi),\label{IFSeries1}
\end{equation}
where $I_{q,\psi_{0}}^{n}$ are the Fourier components of $I_{\psi_{0}}(q,\varphi)$.

For a 2D system of identical particles, the intensity $I_{\psi_{0}}(q,\varphi)$ scattered from a particle in the reference orientation $\psi_{0}=0$
is related to the intensity $I_{\psi_{i}}(q,\varphi)$ scattered from a particle in an arbitrary orientation $\psi_{i}$ as $I_{\psi_{i}}(q,\varphi)=I_{\psi_{0}}(q,\varphi-\psi_{i})$.
Applying the shift theorem for the Fourier transforms \cite{Oppenheim} we obtain for the corresponding Fourier components of the intensities, $I_{q,\psi_{i}}^{n}=I_{q,\psi_{0}}^{n}\exp(-\rmi n\psi_{i})$.
Using these relations we can write for the Fourier components $I_{q}^{n}$ of the intensity $I(q,\varphi)$ scattered from $N$ particles
\begin{equation}
 I_{q}^{n}=I_{q,\psi_{0}}^{n}\sum_{i=1}^{N}\exp(-\rmi n\psi_{i})=I_{q,\psi_{0}}^{n}\mathbf{A}_{n},
\label{FCrelIncoh}
\end{equation}
where $\mathbf{A}_{n}=\sum_{i=1}^{N}\exp(-\rmi n\psi_{i})$ is a random phasor sum \cite{Goodman2}.
According to  (\ref{IFSeries1})
the intensity scattered from a single particle can be uniquely determined by the set of complex coefficients $\{I_{q,\psi_{0}}^{n}\}=\{|I_{q,\psi_{0}}^{n}|,\phi_{q,\psi_{0}}^{n}=\arg(I_{q,\psi_{0}}^{n})\}$.
Here we propose a direct approach for determination of these Fourier components $\{I_{q,\psi_{0}}^{n}\}$ applying two- and three-point CCFs to the measured
intensities $I(q,\varphi)$ scattered form $N$ particles.

We start with the two-point CCF defined at two resolution rings $q_{1}$ and $q_{2}$ \cite{Kam, Saldin, Altarelli}
\begin{eqnarray}
C(q_{1},q_{2},\Delta)&&=\left\langle \widetilde{I}(q_{1},\varphi)\widetilde{I}(q_{2},\varphi+\Delta)\right\rangle_{\varphi},
\label{Cq1q2}
\end{eqnarray}
where $0\leq \Delta \leq2\pi$ is the angular coordinate [see figure~\ref{Fig:CCFScheme}(c)], $\widetilde{I}(q,\varphi)=I(q,\varphi)- \left\langle I(q,\varphi)\right\rangle_{\varphi}$ is the intensity fluctuation function, and $\left\langle \dots \right\rangle_\varphi$ denotes the average over the angle $\varphi$.
It can be directly shown \cite{Altarelli} that the Fourier components $C_{q_{1},q_{2}}^{n}$ of the CCF $C(q_{1},q_{2},\Delta)$ for $n\neq 0$ are defined by the Fourier components $I_{q}^{n}$ of the intensities $I(q,\varphi)$\footnote{We note that the Fourier components of the intensity fluctuation function $\widetilde{I}^{n}_{q}$ can be expressed through the Fourier components of intensity  $I^{n}_{q}$ using relation $\widetilde{I}^{n}_{q} = I^{n}_{q} -I^{0}_{q}\cdot\delta_{n,0}$, where $\delta_{n,0}$ is the Kronecker symbol.}
\begin{equation}
C_{q_{1},q_{2}}^{n}=I_{q_{1}}^{n\ast}\cdot I_{q_{2}}^{n},
\label{FCrelIncoh1}
\end{equation}
and for $n=0$ Fourier components $C_{q_{1},q_{2}}^{n}=0$ according to the definition of the intensity fluctuation function $\widetilde{I}(q,\varphi)$.
Using \eref{FCrelIncoh} in \eref{FCrelIncoh1} and introducing statistical averaging $\left\langle \dots \right\rangle_{M}$ over a large number $M$ of diffraction patterns one can get
\begin{equation}
\left\langle C_{q_{1},q_{2}}^{n}\right\rangle_{M} = I_{q_{1},\psi_{0}}^{n\ast}I_{q_{2},\psi_{0}}^{n} \cdot \left\langle | \mathbf{A}_{n} |^{2}\right\rangle_{M} = I_{q_{1},\psi_{0}}^{n\ast}I_{q_{2},\psi_{0}}^{n} \cdot N.
\label{Cq1q2Iq1Iq2A}
\end{equation}
Here, we used the fact that for a uniform distribution of orientations  of $N$ particles
$\left\langle | \mathbf{A}_{n} |^{2}\right\rangle_{M}$
asymptotically converges to $N$ for a sufficiently large number $M$ of diffraction patterns\footnote{Note, that in this paper, contrary to \cite{Altarelli, Kurta}, we use a not normalized CCFs which, in particular, results in different asymptotic values of $\left\langle | \mathbf{A}_{n} |^{2}\right\rangle_{M}$.}.

\Eref{Cq1q2Iq1Iq2A} can be used to determine both, the amplitudes $|I_{q,\psi_{0}}^{n}|$  and phases $\phi_{q,\psi_{0}}^{n}$ (for $n>0$) of the Fourier components $I_{q,\psi_{0}}^{n}$ associated with a single particle. For example, to determine the amplitudes equation (\ref{Cq1q2Iq1Iq2A}) can be applied successively to three different resolution rings $q_{1},\;q_{2}$ and $q_{3}$ connecting each time a pair of $q$-values. Direct evaluation gives for the amplitudes of the Fourier components of a single particle on the ring $q_{1}$
\begin{equation}
| I_{q_{1},\psi_{0}}^{n}|={\cal I}^n_{q_{1},\psi_{0}} / \sqrt{N},
\label{IqSysEq}
\end{equation}
where ${\cal I}^n_{q_{1},\psi_{0}} = \sqrt{\left|\left\langle C_{q_{1},q_{2}}^{n}\right\rangle_{M}\right|\cdot\left|\left\langle C_{q_{3},q_{1}}^{n}\right\rangle_{M}\right|/\left|\left\langle C_{q_{3},q_{2}}^{n}\right\rangle_{M}\right|}$ is an experimentally determined quantity.
Applying (\ref{IqSysEq}) to different resolution rings $q$ and orders $n$ all required amplitudes $| I_{q,\psi_{0}}^{n}|$ can be determined\footnote{If number of particles in the system is not known {\it a priori} then $N$ in (\ref{IqSysEq}) has to be considered as a scaling factor that is obtained on the final stage of reconstruction of a single particle intensity (see section \ref{sec:recov}).}.
\Eref{IqSysEq} should be used with care to avoid possible instabilities due to division by zero.
For this purpose one should exclude from consideration the cases when $\left|\left\langle C_{q_{3},q_{2}}^{n}\right\rangle_{M}\right|$ is close to zero.
Since the Fourier components of the intensities obey the symmetry condition $I_{q,\psi_{0}}^{n\ast}=I_{q,\psi_{0}}^{-n}$ it is sufficient to determine $I_{q,\psi_{0}}^{n}$ for $n\geq 0$. The  $0$-th order Fourier component by its definition \cite{Altarelli,Kurta} is a real-valued quantity, $I_{q}^{0}=\left\langle I(q,\varphi)\right\rangle_{\varphi}$, and can be determined from the experiment as well. Using (\ref{FCrelIncoh}) one can readily find that $I_{q,\psi_{0}}^{0}=I_{q}^{0}/N$.

\Eref{Cq1q2Iq1Iq2A} also determines the phase difference between two Fourier components $I_{q_{1},\psi_{0}}^{n}$ and $I_{q_{2},\psi_{0}}^{n}$ of the same order $n$, defined at two different resolution rings $q_{1}$ and $q_{2}$,
\begin{equation}
\arg[\left\langle C_{q_{1},q_{2}}^{n}\right\rangle_{M}]=\phi_{q_{2},\psi_{0}}^{n}-\phi_{q_{1},\psi_{0}}^{n}.
\label{PhRel1}
\end{equation}
Notice, that one can freely assign an arbitrary phase to one of the Fourier components $I_{q_{j},\psi_{0}}^{n}$, which corresponds to an arbitrary initial angular orientation of a particle. Assuming, for example, $\phi_{q_{1},\psi_{0}}^{n}=0$ and using (\ref{PhRel1}) one can directly determine the phases of the Fourier components with the same $n$-value on all other resolution rings $q_{2}\neq q_{1}$ \cite{Saldin1}.

To completely solve the phase problem, it is required to obtain additional phase relations between Fourier components with different $n$ values on different resolution rings $q$.
These relations can be determined using a three-point CCF introduced by Kam \cite{Kam, Kam1}.
The important aspect of our approach is to use the three-point CCF defined on {\it three} different resolution rings [see figure~\ref{Fig:CCFScheme}(d)],
\begin{equation}
C(q_{1},q_{2},q_{3},\Delta_{1},\Delta_{2})=\left\langle \widetilde{I}(q_{1},\varphi)\widetilde{I}(q_{2},\varphi+\Delta_{1})\widetilde{I}(q_{3},\varphi+\Delta_{2})\right\rangle_{\varphi},
\label{Cq1q2q3}
\end{equation}
contrary to \cite{Saldin, Saldin1} where the three-point CCF was defined on {\it two} resolution rings.
Similar to $C_{q_{1},q_{2}}^{n}$ it can be shown (see Appendix) that for the Fourier components of this CCF the following relation is valid,
\begin{equation}
C_{q_{1},q_{2},q_{3}}^{n_{1},n_{2}}=I_{q_{1}}^{(n_{1}+n_{2})\ast} I_{q_{2}}^{n_{1}} I_{q_{3}}^{n_{2}},
\label{Cq1q2q3FT1}
\end{equation}
for $n_{1}\neq 0,n_{2}\neq 0,n_{1}\neq -n_{2}$.
Using (\ref{FCrelIncoh}) in (\ref{Cq1q2q3FT1}) and performing statistical averaging we obtain for the averaged Fourier components of the three-point CCF
\begin{equation}
\left\langle C_{q_{1},q_{2},q_{3}}^{n_{1},n_{2}}\right\rangle_{M}=I_{q_{1},\psi_{0}}^{(n_{1}+n_{2})\ast} I_{q_{2},\psi_{0}}^{n_{1}} I_{q_{3},\psi_{0}}^{n_{2}}
\cdot \left\langle \mathbf{A}_{n_{1},n_{2}} \right\rangle_{M},
\label{Cq1q2q3FT2}
\end{equation}
where $\mathbf{A}_{n_{1},n_{2}}=\sum_{i,j,k=1}^{N}\exp\{\rmi[(n_{1}+n_{2})\psi_{i}-n_{1}\psi_{j}-n_{2}\psi_{k}]\}$.
Our analysis shows (see Appendix) that
the statistical average $\left\langle \mathbf{A}_{n_{1},n_{2}} \right\rangle_{M}$ converges to $N$ for a sufficiently large number $M$ of diffraction patterns,
i.e. $\arg[\left\langle \mathbf{A}_{n_{1},n_{2}} \right\rangle_{M}]=0$, and we get from (\ref{Cq1q2q3FT2}) the following phase relation
\begin{equation}
\arg[\left\langle C_{q_{1},q_{2},q_{3}}^{n_{1},n_{2}} \right\rangle_{M}]=\phi_{q_{2},\psi_{0}}^{n_{1}}+\phi_{q_{3},\psi_{0}}^{n_{2}}-\phi_{q_{1},\psi_{0}}^{(n_{1}+n_{2})}.
\label{PhRel2}
\end{equation}
\Eref{PhRel2} determines the phase shift between three Fourier components $I_{q_{1},\psi_{0}}^{(n_{1}+n_{2})}$, $I_{q_{2},\psi_{0}}^{n_{1}}$, and $I_{q_{3},\psi_{0}}^{n_{2}}$ of different order $n$ defined on three resolution rings. If $n_{1}=n_{2}=n$ and $n_{3}=2n$, equation (\ref{PhRel2}) reduces to a particular form, giving the phase relation between Fourier components of only two different orders $n$ and $2n$,
\begin{equation}
\arg[\left\langle C_{q_{1},q_{2},q_{3}}^{n,n}\right\rangle_{M}]=\phi_{q_{2},\psi_{0}}^{n}+\phi_{q_{3},\psi_{0}}^{n}-\phi_{q_{1},\psi_{0}}^{2n}.
\label{PhRel2a}
\end{equation}

Equations~(\ref{IqSysEq}), (\ref{PhRel1}) and (\ref{PhRel2}) constitute the core of our approach and allow us to
directly and unambiguously determine the complex Fourier components $I_{q,\psi_{0}}^{n}$ using measured x-ray data from a disordered system of $N$ particles.
The obtained Fourier components $I_{q,\psi_{0}}^{n}$ can be used in  (\ref{IFSeries1}) to recover the scattered intensity $I_{\psi_{0}}(q,\varphi)$ corresponding to a single particle.

\section{Recovery of the projected electron density of a single particle \label{sec:recov}}

\begin{figure}[!htbp]
\centering
\includegraphics[width=0.48\textwidth]{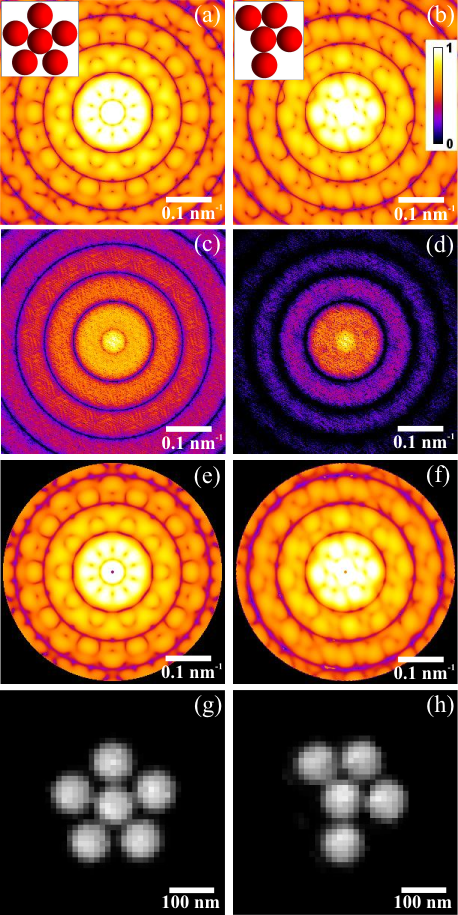}
\caption{\label{Fig:Results}(Color online)
(a),(b) Scattered intensity (logarithmic scale) calculated for a single pentagonal cluster (a) and an asymmetric cluster (b) (clusters are shown in the insets).
(c),(d) Coherently scattered intensity from a disordered system consisting of $N=10$ clusters in random position and orientation. Scattered signal corresponding to the incident fluence of $10^{12}$  and $10^{13}\;\rm{photons}/25\,\mu\rm{m}^2$ for pentagonal and asymmetric clusters correspondingly. Diffraction pattern shown in (d) also contains Poisson noise. (e),(f) Scattered intensity corresponding to a single pentagonal (e) and asymmetric (f) clusters recovered from $M=10^5$ diffraction patterns of the form (c) and (d) correspondingly.
(g),(h) Structure of a single cluster reconstructed by an iterative phase retrieval algorithm using the diffraction patterns shown in (e) and (f).}
\end{figure}

We demonstrate our approach for the case of a coherent illumination of a disordered system of particles, in the presence of Poisson noise in the scattered signal.
We recover diffraction patterns and projected electron densities for two different particles, a centered pentagonal cluster [figure~\ref{Fig:Results}(a)] that has 5-fold rotational symmetry and an asymmetric cluster [figure~\ref{Fig:Results}(b)]. Both clusters have a size of $d=300\;\rm{nm}$ and are composed of polymethylmethacrylate (PMMA) spheres of $50\;\rm{nm}$ radius.

In both cases we consider kinematical coherent scattering of x-rays with wavelength $\lambda=1\;\mathring{A}$ from a system of $N=10$ clusters in random positions and orientations, distributed within a sample area of $5\times 5 \;\mu\rm{m}^{2}$ [see figure~\ref{Fig:CCFScheme}(a)].
Diffraction patterns are simulated for a 2D detector of size $D=24\;\rm{mm}$
(with pixel size $p=80\;\mu\rm{m}$), positioned in the transmission geometry at $L=3\;\rm{m}$ distance from the sample [see figure~\ref{Fig:ExpGeometr}]. This experimental geometry corresponds to scattering to a maximum resolution of 0.25 nm$^{-1}$.
For given experimental conditions the speckle size corresponding to the illuminated area is below the pixel size of the detector. At the same time the speckle size corresponding to the size of a single particle is
about $12$ pixels, that provides sufficient sampling for phase the retrieval.

The coherently scattered intensities simulated for single realizations of the systems\footnote{All simulations of diffraction patterns were performed using the computer code MOLTRANS.} are shown in figures~\ref{Fig:Results}(c) and \ref{Fig:Results}(d). We note, that in simulations the intensity $I(q,\varphi)$ scattered from a system of $N$ particles was calculated as a coherent sum of the scattered amplitudes $A_{\psi_{i}}(q,\varphi)$ from each particle, i.e, $I(q,\varphi)=|\sum_{i=1}^{N}A_{\psi_{i}}(q,\varphi)|^2$. The incident fluence was considered to be $10^{12}$ and $10^{13}\;\rm{photons}/25\,\mu\rm{m}^2$ for pentagonal and asymmetric clusters, respectively. In the case of asymmetric clusters additional Poisson noise was included in the simulations. The Fourier components of two-point and three-point CCFs [equations (\ref{Cq1q2Iq1Iq2A}) and (\ref{Cq1q2q3FT2})] were averaged over $M=10^5$ diffraction patterns\footnote{According to our simulations, two-point and three-point CCFs defined on different resolution rings $q_{1}$, $q_{2}$ and $q_{3}$ [see (\ref{Cq1q2}) and (\ref{Cq1q2q3})] even in the case of coherent illumination of a disordered system do not contain the inter-particle contribution. Contrary to that, as it was shown in \cite{Altarelli,Kurta}, this is not the case for two-point CCFs defined on the same resolution ring $q_{1}=q_{2}$, when inter-particle contributions and noise can give a substantial contribution. Therefore, in our present work we avoided this last case for the unique determination of amplitudes and phases of the Fourier coefficients of single particles.}.

Once all amplitudes $\left\vert I_{q, \psi_{0}}^{n}\right\vert$ and phases $\phi_{q,\psi_{0}}^{n}$ of the Fourier components $I_{q, \psi_{0}}^{n}$ are determined, the diffraction pattern corresponding to a single particle can be recovered by performing the Fourier transform \eref{IFSeries1}.
Using experimentally accessible quantities $\left\langle \left\langle I(q,\varphi)\right\rangle_{\varphi}\right\rangle_{M}=NI_{q,\psi_{0}}^{0}$ and
${\cal I}^{n}_{q,\psi_{0}}=\sqrt{N}\left\vert I_{q,\psi_{0}}^{n}\right\vert$, we can rewrite \eref{IFSeries1} as
\begin{equation}
I_{\psi_{0}}(q,\varphi)=\frac{\left\langle\left\langle I(q,\varphi)\right\rangle_{\varphi}\right\rangle_{M}}{N} +\sum_{n=-\infty\atop n\neq 0}^{\infty}
\frac{{\cal I}^{n}_{q,\psi_{0}}}{\sqrt{N}}\exp(\rmi\phi_{q,\psi_{0}}^{n})\exp(\rmi n\varphi).\label{IFSeries3}
\end{equation}

\begin{figure}[!htbp]
\centering
\includegraphics[width=1.0\textwidth]{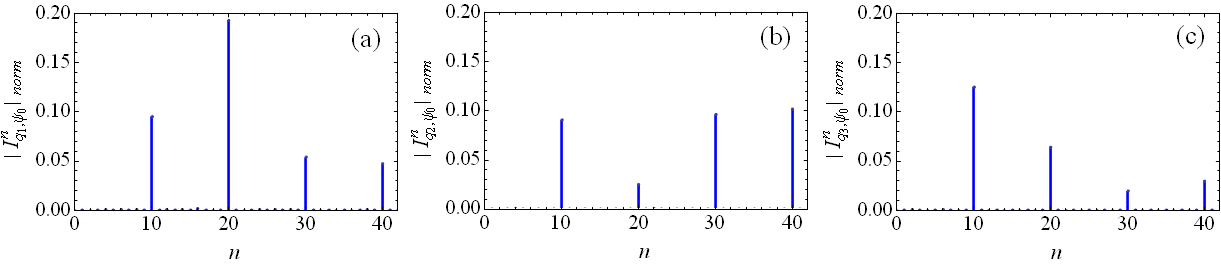}
\caption{\label{Fig:Spectra1}(Color online) Normalized amplitudes $|I_{q,\psi_{0}}^{n}|_{norm}=|I_{q,\psi_{0}}^{n}|/(|I_{q,\psi_{0}}^{0}|\sqrt{N})$ determined using (\ref{IqSysEq}) at three different resolution rings  (a) $q_{1}=0.21\;\rm{nm}^{-1}$, (b) $q_{2}=0.23\;\rm{nm}^{-1}$ and (c) $q_{3}=0.25\;\rm{nm}^{-1}$.}
\end{figure}

Here we discuss determination of the amplitudes and phases of the Fourier components $I_{q,\psi_{0}}^{n}$ associated with a single particle,
using x-ray data from 2D system of pentagonal clusters [figure~\ref{Fig:Results}(a)] described above.
Applying expression (\ref{IqSysEq}) to different $q$ we find all required amplitudes $\sqrt{N}|I_{q,\psi_{0}}^{n}|$ (for $n>0$) scaled by the factor $\sqrt{N}$.
The $0$-th order Fourier component is determined as $I_{q,\psi_{0}}^{0}=\left\langle \left\langle I(q,\varphi)\right\rangle_{\varphi}\right\rangle_{M}/N$.
In figure~\ref{Fig:Spectra1} the values $\sqrt{N}|I_{q,\psi_{0}}^{n}|$ derived from (\ref{IqSysEq}) normalized by $\left\langle\left\langle I(q,\varphi)\right\rangle_{\varphi}\right\rangle_{M}$ are shown for $n\le40$  at three different resolution rings $q_{1}=0.21\;\rm{nm}^{-1}$, $q_{2}=0.23\;\rm{nm}^{-1}$ and $q_{3}=0.25\;\rm{nm}^{-1}$.
Due to 5-fold symmetry of the pentagonal cluster, in the $q$-range accessible in our model experiment we need to determine the phases
of the Fourier components $I_{q,\psi_{0}}^{n}$ only with $n=10\cdot l,\;1\leq l\leq 4$, where $l$ is an integer.
Below we present the algorithm for the phase determination using equations (\ref{PhRel1}), (\ref{PhRel2}) and (\ref{PhRel2a}).

\begin{enumerate}
\item First, we assign an arbitrary phase to one of the Fourier components $I_{q,\psi_{0}}^{n}$, which corresponds to an arbitrary initial angular orientation of a particle.
It is convenient to assign this phase to the Fourier component with the smallest available $n$ value. In our case we assign a zero phase $\phi_{q_{1},\psi_{0}}^{10}=0$ to the Fourier component $I_{q_{1},\psi_{0}}^{10}$.
\item The phases of the Fourier components with $n=10$ on other resolution rings $q_{j}\;(j=1,2)$ were obtained by using (\ref{PhRel1}), i.e., $\phi_{q_{j},\psi_{0}}^{10}=\phi_{q_{1},\psi_{0}}^{10}+\arg[\left\langle C_{q_{1},q_{j}}^{10}\right\rangle_{M}]=\arg[\left\langle C_{q_{1},q_{j}}^{10}\right\rangle_{M}]$.
\item The phase of the Fourier component $I_{q_{1},\psi_{0}}^{20}$ on the resolution ring $q_{1}$ was determined by using (\ref{PhRel2a}) with $n=10$,
$\phi_{q_{1},\psi_{0}}^{20}=\phi_{q_{2},\psi_{0}}^{10}+\phi_{q_{3},\psi_{0}}^{10}-\arg[\left\langle C_{q_{1},q_{2},q_{3}}^{10,10}\right\rangle_{M}]$.
\item The phases of the Fourier components with $n=20$ on other resolution rings $q_{j}\;(j=1,2)$ were obtained from (\ref{PhRel1}) $\phi_{q_{j},\psi_{0}}^{20}=\phi_{q_{1},\psi_{0}}^{20}+\arg[\left\langle C_{q_{1},q_{j}}^{20}\right\rangle_{M}]$.
\item The phase of the Fourier component $I_{q_{1},\psi_{0}}^{30}$ on the resolution ring $q_{1}$ was determined
by using (\ref{PhRel2}) with $n_{1}=10$ and $n_{2}=20$,
$\phi_{q_{1},\psi_{0}}^{30}=\phi_{q_{2},\psi_{0}}^{10}+\phi_{q_{3},\psi_{0}}^{20}-\arg[\left\langle C_{q_{1},q_{2},q_{3}}^{10,20}\right\rangle_{M}]$.
\end{enumerate}

The process was continued in a similar way until all phases at each $q$-value were determined.
The same procedure has been applied for the case of 2D system of asymmetric clusters [figure~\ref{Fig:Results}(b)]. Due to asymmetric structure of particles, in the same $q$-range
we need to determine the phases of a significantly larger set of the Fourier components with $n=2\cdot k,\;1\leq k\leq24$, where $k$ is an integer. Assigning a zero phase $\phi_{q_{1},\psi_{0}}^{2}=0$
to the Fourier component with $n=2$ we successively determine the phases of the Fourier components of the higher orders up to $n=48$.
We note, that the data redundancy intrinsic to $\left\langle C_{q_{1},q_{2}}^{n}\right\rangle_{M}$ and $\left\langle C_{q_{1},q_{2},q_{3}}^{n_{1},n_{2}}\right\rangle_{M}$
offers a lot of flexibility in determination the possible ways of solving the phases of the Fourier components $I_{q,\psi_{0}}^{n}$.

\begin{figure}[!ht]
\centering
\includegraphics[width=0.45\textwidth]{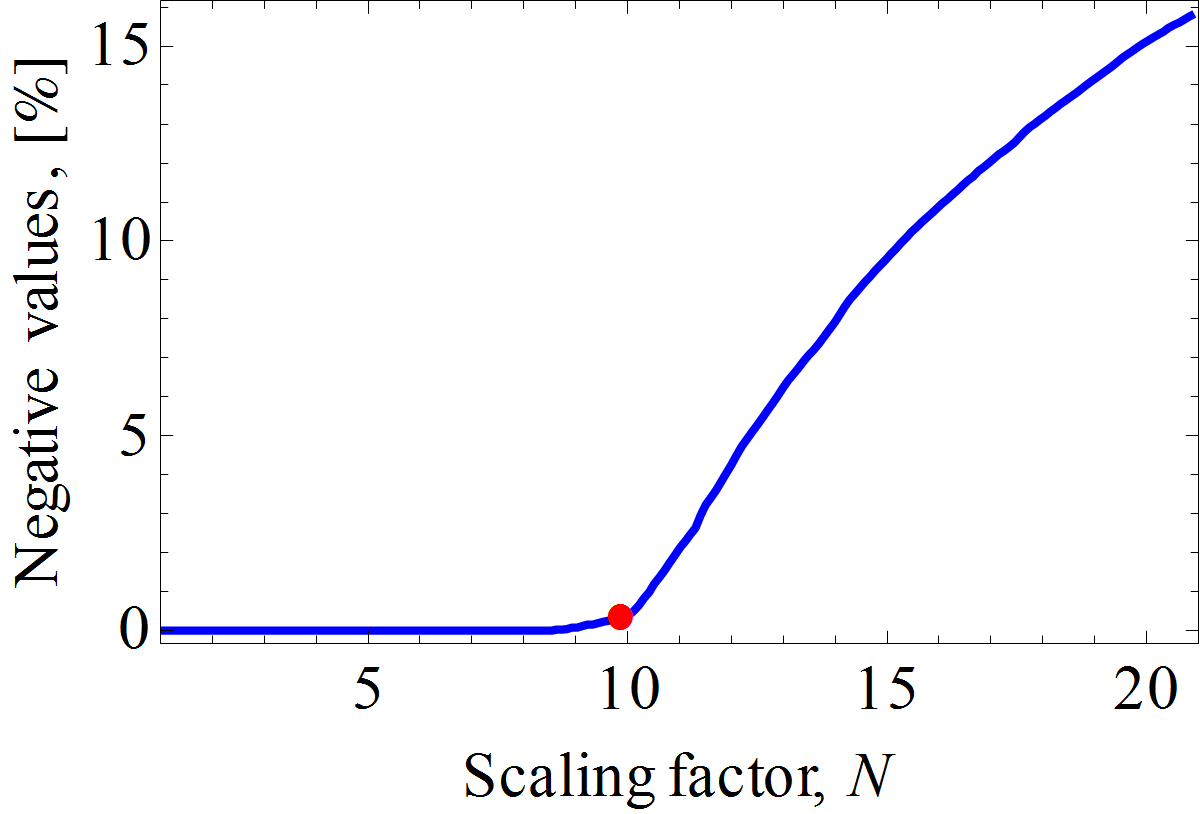}
\caption{\label{Fig:ScalingZ}(Color online) Dependence of the relative number of pixels with negative values (normalized to the total number of pixels on the detector) in the recovered diffraction pattern $I_{\psi_{0}}(q,\varphi)$ as a function of the scaling factor $N$. The red point indicates the value of $N=9.8$ used in the final reconstruction of intensity $I_{\psi_{0}}(q,\varphi)$.}
\end{figure}

If the number of particles $N$ in the system is not known [see equation~\eref{IFSeries3}], it can be determined using the positivity constraint $I_{\psi_{0}}(q,\varphi)\geq 0$ for the recovered intensity.
One can slightly relax this constraint and allow a small fraction of pixels with negative values in order to get an image with a better contrast. These negative values can be substituted by zeros before using this diffraction pattern in the iterative phase retrieval algorithms for the reconstruction of the particle structure.  This procedure can be justified by inaccuracies which arise due to statistical estimate of the CCF and also different errors in calculations. For example, for the case of pentagonal clusters in our simulations we applied the value of $N=9.8$, which corresponds to $0.27 \%$ of negative pixes on the detector [see figure~\ref{Fig:ScalingZ}]. This value is very close to the theoretically predicted $N=10$ for a uniform distribution of orientations of particles.

\section{Results and discussion}

The diffraction patterns corresponding to single particles recovered by our approach are presented in figures~\ref{Fig:Results}(e) and \ref{Fig:Results}(f). As one can see from these figures, the recovered single particle diffraction patterns reproduce the diffraction patterns of individual clusters shown in figures~\ref{Fig:Results}(a) and ~\ref{Fig:Results}(b) very well. The structure of the clusters reconstructed from these recovered diffraction patterns was obtained by a standard phase retrieval approach and is presented in figures~\ref{Fig:Results}(g) and \ref{Fig:Results}(h). In our reconstructions we used an alternative sequence of hybrid input-output (HIO) and error-reduction (ER) algorithms \cite{Fienup}.
The comparison of the single cluster structures obtained by our approach [figures~\ref{Fig:Results}(g) and \ref{Fig:Results}(h)] with the initial model shown in the insets of figures~\ref{Fig:Results}(a) and \ref{Fig:Results}(b) confirms the correctness of our reconstruction. Both clusters were reconstructed with a resolution up to $25\;\rm{nm}$. These results clearly demonstrate the ability of our approach to recover the single-particle structure from noisy data obtained in coherent x-ray scattering experiments.

The ability of the presented approach to recover the diffraction pattern of a single particle relays on the accuracy of the two- and three- point CCFs determination. The statistical properties of
the CCFs and their convergence to the average values strongly depend on the number of particles $N$ in the system, their density and the distribution of their orientations \cite{Altarelli, Kurta}.
In particular, the value of the scaling factor $N$ in the expression of the Fourier components of the CCF is defined by the statistical distribution of particles in the system. We performed a few simulations with varying number $N$ of particles in the system: a) for a Gaussian distribution of the number of particles with an average value $\langle N \rangle=20$ and standard deviation $\sigma=2$; b) for a Poisson distribution with an average number of particles $\langle N \rangle=30$. In both cases the CCFs statistically converge to the same average values as for the systems with a fixed number of particles, $N=20$ and $N=30$ (for Gaussian and Poisson distributions, correspondingly). The only difference we observed is a slightly slower convergence of CCFs in the case of statistical distribution of particles compared to the case with a fixed number $N$. This means, that in the case of systems with a varying number of particles one needs to measure a larger number of diffraction patterns.

Particle density also affects the statistical behaviour of the CCFs. In the limiting case of a very dilute system the CCFs calculated for a single diffraction pattern are defined by
independent structural contributions of $N$ individual particles. In this case statistical averaging over diffraction patterns is, in fact, acts on the fluctuating terms $|\mathbf{A}_{n} |^{2}$ in \eref{Cq1q2Iq1Iq2A} and $\mathbf{A}_{n_{1},n_{2}}$ in \eref{Cq1q2q3FT2} and convergence of these terms to their statistical estimates determines the number $M$ of diffraction patterns required for averaging \cite{Kurta}. In the case of a dense system, the CCFs contain a significant inter-particle contribution in the same range of $q$-values, where the structural contribution of individual particles is observed \cite{Altarelli, Kurta}. This additional contribution slows down the convergence of CCFs (especially of the three-point CCF) and the number of measured diffraction patterns should be increased. In the presence of Poisson noise in the scattered signal one needs to accumulate even more diffraction patterns. Additional effects of resolution, solution scattering, {\it etc.} on the measured CCFs were considered in detail in \cite{Kirian}.
Results of our simulations with PMMA clusters and asymptotic estimates presented in \cite{Kurta} show that, in practice, the number of particles in the system should be less than few dozens.
This allows to achieve statistical convergence of the CCFs using $10^5-10^6$ diffraction patterns. In the disordered systems considered in this work the particle density is characterized by $\langle R\rangle/d\approx5.0$, where $\langle R\rangle$ is the average inter-particle distance. At these conditions it was sufficient to use $10^5$ diffraction patterns with Poisson noise to achieve convergence of the two and three-point CCFs and perform successful reconstruction of the particle structure. In practical applications the convergence of the CCFs can be directly controlled as a function of the number $M$ of the diffraction patterns considered in the averaging \cite{Kurta}.

\section{Conclusions}

In conclusion, we present here a direct, non-iterative approach for the unambiguous recovery of the scattering pattern corresponding to a single particle (molecule, cluster, etc) using scattering data from a disordered system of these particles including noise. Our simulations demonstrate the successful application of this approach to 2D systems composed of particles with and without rotational symmetry. It can be of particular interest for studies of membrane proteins which naturally form 2D systems. Our approach can be also applied to 3D systems of particles to recover their projected  electron density, if a specific alignment of the particles along the direction of the incoming x-ray beam can be achieved. We have shown that our approach is robust to noise and intensity fluctuations which arise due to coherent interference of waves scattered from different particles. We foresee that this method will find a wide application in the studies of disordered systems at newly emerging free-electron lasers.

\ack

We acknowledge a careful reading of the manuscript by H. Franz.
Part of this work was supported by BMBF Proposal No. 05K10CHG ``Coherent Diffraction Imaging and Scattering of Ultrashort Coherent Pulses with Matter'' in the
framework of the German-Russian collaboration ``Development and Use of Accelerator-Based Photon Sources'' and the Virtual Institute VH-VI-403 of the Helmholtz association.

\appendix
\section*{Appendix}
\setcounter{section}{1}

The Fourier series expansion of the CCF  \eref{Cq1q2q3} can be written as
\begin{eqnarray}
\fl C(q_{1},q_{2},q_{3},\Delta_{1},\Delta_{2})&&=
\sum_{n_{1}=-\infty\atop n_{1}\neq 0}^{\infty}
\sum_{n_{2}=-\infty\atop n_{2}\neq 0}^{\infty}
 C_{q_{1},q_{2},q_{3}}^{n_{1},n_{2}}e^{in_{1}\Delta_{1}}e^{in_{2}\Delta_{2}},\quad n_{1}\neq -n_{2}, \label{Cq1q2q3FT1general1}\\
C_{q_{1},q_{2},q_{3}}^{n_{1},n_{2}}&&=\left(\frac{1}{2\pi}\right)^{2}\int_{0}^{2\pi}\int_{0}^{2\pi}C(q_{1},q_{2},q_{3},\Delta_{1},\Delta_{2})\rme^{-\rmi n_{1}\Delta_{1}}\rme^{-\rmi n_{2}\Delta_{2}}d\Delta_{1}d\Delta_{2}\nonumber\\
&&=I_{q_{1}}^{(n_{1}+n_{2})\ast}I_{q_{2}}^{n_{1}}I_{q_{3}}^{n_{2}},\quad n_{1}\neq 0,n_{2}\neq 0,n_{1}\neq -n_{2}.\label{Cq1q2q3FT1general}
\end{eqnarray}
where $C_{q_{1},q_{2},q_{3}}^{n_{1},n_{2}}$ are the Fourier components of the CCF $C(q_{1},q_{2},q_{3},\Delta_{1},\Delta_{2})$.
In general, (\ref{Cq1q2q3FT1general}) determines a relation between three different Fourier components of intensity $I_{q}^{n}$ of the order $n_{1}$, $n_{2}$ and $n_{1}+n_{2}$, defined on three resolution rings, $q_{1},q_{2}$ and $q_{3}$.

Considering incoherent scattering from $N$ particles, we can rewrite (\ref{Cq1q2q3FT1general}) in terms of the Fourier components of intensity $I_{q,\psi_{0}}^{n}$ associated with a single particle,
\begin{equation}
\left\langle C_{q_{1},q_{2},q_{3}}^{n_{1},n_{2}}\right\rangle_{M}=I_{q_{1},\psi_{0}}^{(n_{1}+n_{2})\ast}I_{q_{2},\psi_{0}}^{n_{1}}I_{q_{3},\psi_{0}}^{n_{2}}
\cdot \left\langle \mathbf{A}_{n_{1},n_{2}} \right\rangle_{M},
\label{Cq1q2q3FT2a}
\end{equation}
where $\left\langle\dots \right\rangle_{M}$ denotes the statistical averaging and we introduce a new random phasor sum
\begin{equation}
\mathbf{A}_{n_{1},n_{2}}=\sum_{i,j,k=1}^{N}\exp\{\rmi[(n_{1}+n_{2})\psi_{i}-n_{1}\psi_{j}-n_{2}\psi_{k}]\}.
\label{An1n2}
\end{equation}
To understand the statistical properties of $\left\langle \mathbf{A}_{n_{1},n_{2}} \right\rangle_{M}$
we split it into three contributions,
\begin{equation}
\left\langle \mathbf{A}_{n_{1},n_{2}} \right\rangle_{M}=\left\langle \mathbf{B}_{n_{1},n_{2}} \right\rangle_{M}+\left\langle \mathbf{C}_{n_{1},n_{2}} \right\rangle_{M}+\left\langle \mathbf{D}_{n_{1},n_{2}} \right\rangle_{M},
\label{ASumB}
\end{equation}
where $\left\langle \mathbf{B}_{n_{1},n_{2}} \right\rangle_{M}$, $\left\langle \mathbf{C}_{n_{1},n_{2}} \right\rangle_{M}$ and $\left\langle \mathbf{D}_{n_{1},n_{2}} \right\rangle_{M}$ are defined for different combinations of the subscripts $i, j$ and $k$,
\begin{eqnarray}
\left\langle \mathbf{B}_{n_{1},n_{2}} \right\rangle_{M}&&=\left\langle \sum_{i=j=k}\exp\{\rmi[(n_{1}+n_{2})\psi_{i}-n_{1}\psi_{j}-n_{2}\psi_{k}]\}\right\rangle_{M}=N,
\label{B1a}\\
\left\langle \mathbf{C}_{n_{1},n_{2}} \right\rangle_{M}&&=\left\langle \sum_{i=j,j\neq k}\dots + \sum_{i=k,j\neq k}\dots +\sum_{i\neq j,j=k}\dots\right\rangle_{M}\nonumber\\
&&=2\left\langle \sum_{i>j}\cos[n_{1}(\psi_{i}-\psi_{j})]\right\rangle_{M}+2\left\langle\sum_{i>j}\cos[n_{2}(\psi_{i}-\psi_{j})]\right\rangle_{M}\nonumber\\
&&+2\left\langle\sum_{i>j}\cos[(n_{1}+n_{2})(\psi_{i}-\psi_{j})]\right\rangle_{M},\label{B3a}\\
\left\langle \mathbf{D}_{n_{1},n_{2}} \right\rangle_{M}&&=\left\langle \sum_{i\neq j\neq k}\exp\{\rmi[(n_{1}+n_{2})\psi_{i}-n_{1}\psi_{j}-n_{2}\psi_{k}]\}\right\rangle_{M}.\label{B2a}
\end{eqnarray}
Now we determine the statistical estimates of the terms $\left\langle \mathbf{C}_{n_{1},n_{2}} \right\rangle_{M}$ and  $\left\langle \mathbf{D}_{n_{1},n_{2}} \right\rangle_{M}$. We consider an uniform distribution of orientations of particles $\psi$ on the interval $(-\pi, \pi)$, where two different orientations $\psi_{i}$ and $\psi_{j}$ are independent and equally probable. We also consider that the sum (or difference) of two angles $\psi$, as well as a product $n\psi$ (where $n$ is an arbitrary number) is also distributed on the interval $(-\pi, \pi)$. Another words, we deal with a wrapped angular distribution \cite{CircStat1}. In this case, we can use two following arguments. First, the difference $\bar{\psi}=\psi_{i}-\psi_{j}$ of two uniformly distributed independent variables  $\psi_{i}$ and $\psi_{j}$ also has a uniform distribution.
In this case the probability density function (PDF) of $\bar{\psi}$ is a convolution of the corresponding PDFs of $\psi_{i}$ and $\psi_{j}$ \cite{Goodman1}, which is just a constant after wrapping to the interval $(-\pi, \pi)$. Second, using the rules of probability theory for transformation of random variables \cite{Goodman1,Rohatgi}, it can be shown that the product $n\psi$ is also a uniformly distributed variable.
Using a combination of these two arguments in (\ref{B3a}) and (\ref{B2a}) we find that $\left\langle \mathbf{C}_{n_{1},n_{2}} \right\rangle_{M}=0$ and $\left\langle \mathbf{D}_{n_{1},n_{2}} \right\rangle_{M}=0$. Therefore, the statistical estimate in (\ref{ASumB}) becomes a real number, $\left\langle \mathbf{A}_{n_{1},n_{2}} \right\rangle_{M}=N$. Using this result in (\ref{Cq1q2q3FT2a}) we obtain \eref{PhRel2}.



\end{document}